\documentclass[a4paper,12pt]{article}

\usepackage{epsfig,subfig}
\usepackage{amssymb,amsmath}

\newcommand{\lapprox}{%
\mathrel{%
\setbox0=\hbox{$<$}
\raise0.6ex\copy0\kern-\wd0
\lower0.65ex\hbox{$\sim$}
}}
\newcommand{\gapprox}{%
\mathrel{%
\setbox0=\hbox{$>$}
\raise0.6ex\copy0\kern-\wd0
\lower0.65ex\hbox{$\sim$}
}}

\newcommand{\ba}{\begin{array}}
\newcommand{\ea}{\end{array}}
\newcommand{\bd}{\begin{displaymath}}
\newcommand{\ed}{\end{displaymath}}
\newcommand{\beq}{\begin{equation}}
\newcommand{\eeq}{\end{equation}}
\newcommand{\bea}{\begin{eqnarray}}
\newcommand{\eea}{\end{eqnarray}}
\newcommand{\Z}{\mathbb{Z}}

\newcommand{\ra}{\rightarrow}
\def\q2 {q^2}

\def\bt{\begin{table}}
\def\et{\end{table}}

\catcode`@=11 
\def \gsim{\mathrel{\mathpalette\@versim>}}
\def \lsim{\mathrel{\mathpalette\@versim<}}
\def \@versim#1#2{\lower0.4ex\vbox{\baselineskip\z@skip\lineskip\z@skip
     \lineskiplimit\z@\ialign{$\m@th#1\hfil##\hfil$%
     \crcr#2\crcr\sim\crcr}}}
\catcode`@=12 

\begin{document}

\begin{flushright}
{\small RECAPP-HRI-2015-002 }
\end{flushright}

\begin{center}

{\large\bf Dark matter candidate in an extended
type III seesaw scenario }\\[15mm] 
Avinanda Chaudhuri $^{a,}$\footnote{E-mail: avinanda@hri.res.in},
Najimuddin Khan $^{b,}$\footnote{E-mail: phd11125102@iiti.ac.in},
Biswarup Mukhopadhyaya $^{a,}$\footnote{E-mail: biswarup@hri.res.in} 
and 
Subhendu Rakshit $^{b,}$\footnote{E-mail: rakshit@iiti.ac.in}
\\[2mm]
{\em $^a$Regional Centre for Accelerator-based Particle Physics \\
     Harish-Chandra Research Institute\\
Chhatnag Road, Jhusi, Allahabad - 211~019, India}
\\[2mm]
{\em $^b$Discipline of Physics, Indian Institute of Technology Indore\\
     IET-DAVV Campus, Indore - 452~017, India }\\
[20mm] 
\end{center}

\begin{abstract} 
The type~III seesaw mechanism for neutrino mass generation usually
makes use of at least two $Y = 0$, $SU(2)_L$ lepton triplets. We augment such a model with a third triplet and a sterile neutrino, both of which are odd under a conserved $\Z_2$ symmetry. With all new physics confined to the $\Z_2$-odd sector, whose low energy manifestation is in some higher-dimensional operators, a fermionic dark matter candidate is found to emerge. We identify the region of the parameter space of the scenario, which is consistent with all constraints from relic density and direct searches, and allows a wide range of masses for the dark matter candidate.

\end{abstract}

\vskip 1 true cm

\newpage
\setcounter{footnote}{0}

\def\baselinestretch{1.5}
\section{Introduction}
The standard model (SM) of particle physics has been extremely successful in explaining almost all experimental results. However, extension of the SM is widely expected, to address some yet unexplained observations, which include tiny  masses and the mixing pattern of neutrinos and existence of dark matter (DM) in the Universe.

One possible explanation of small neutrino masses can be obtained by adding $Y = 0$  $SU(2)_L$ lepton triplets to the SM. The active neutrino mixes with the neutral member of such a fermion triplet, allowing $\Delta L=2$ Majorana mass terms for the neutrinos. The heavy Majorana mass of the  neutral member of the fermion triplet ensures smallness of the neutrino mass {\it via} type-III seesaw mechanism. As neutrino oscillation experiments provide information only on neutrino mass-squared differences, the observed data can be explained if at least two neutrinos are massive. Hence neutrino masses in the type-III seesaw mechanism can be explained if we have at least two fermion triplets.

On the other hand, there is strong empirical evidence for the existence of dark matter in the Universe. A simple and attractive candidate for DM should be a new elementary particle that is electrically neutral and stable on cosmological scales. In this work, we would like to explore if the neutral member of a fermion triplet can provide us an answer in this regard. 

Our proposal is an extension of the type-III seesaw model for neutrino masses. As two fermion triplets are required to generate neutrino masses, we need to introduce a third triplet to take care of the DM candidate. This triplet needs to be protected by a $\Z_2$ symmetry to ensure stability of the DM particle. The $SU(2)$ symmetry will in general demand the charged and neutral components of the fermion triplet to be degenerate. However, we need a mass splitting between the charged and neutral component to obtain a DM particle of mass of the order of the electroweak scale. In some models this mass splitting between the charged and neutral components of the fermion triplet have been obtained through radiative correction. Refs.~\cite{Ma:2008cu} and \cite{Chao:2012sz} discussed TeV scale DM in the context of fermion triplets using radiative correction. 
The mass of the charged member of the triplet in such a case has to be well above $\sim$ TeV, in order to avoid fast $t$-channel annihilation and a consequent depletion in relic density. As a result, such models end up with somewhat inflexible prediction of DM mass in the region $2.5-2.7$  TeV.

We propose, as an alternative, two $\Z_2$-odd fermion fields - a $Y = 0$ triplet and a neutral singlet that appears as a heavy sterile neutrino. All SM fields and the first two fermion triplets are even  under this $\Z_2$. Mixing between neutral component of the triplet and the sterile neutrino can yield a $\Z_2$ odd neutral fermion state that is lighter than all other states in the $\Z_2$ odd sector. This fermionic state emerges as the DM candidate in this work. Furthermore, the requisite rate of annihilation is ensured by postulating some $\Z_2$ preserving dimension-five operators. These operators, together with the fact that the neutral component of the DM candidate has $W$ couplings, allow a larger region of the parameter space than what we would have had with a sterile neutrino alone. This interplay brings an enriched DM phenomenology compared to models with only singlet or triplets. This we feel is a desirable feature of our model given the contradictory claims on the mass of a DM candidate from direct as well as indirect searches. Although choice of a light sterile neutrino can lead to light DM candidates, in our model, we will work with DM masses of more than 62.5~GeV to avoid an unacceptably large invisible Higgs decay width.

In section 2, we present the theoretical framework of our model and introduce the mixing term between $\Z_2$ odd triplet $\Sigma$ and sterile neutrino $\nu_s$. The details of our findings i.e. the dependence of relic density and dark matter mass on the various parameters used in the model are presented in section 3. We summarize and conclude in section 4.

\section{Theoretical Framework}
\label{S:theory}
In this section we give a description of our model. We have added three fermion triplets to the SM Lagrangian. The fermion triplets 
$\Sigma = (\Sigma^+, \Sigma^0, \Sigma^-)$ are represented by
the $2 \times 2$ matrix
\bea\label{triplet}
\Sigma &=& 
\left(
\begin{array}{cc}
   \Sigma^0/\sqrt{2}  &   \Sigma^+ \\
     \Sigma^- &  -\Sigma^0/\sqrt{2} 
\end{array}\right),
\eea
Two of these triplets, which are even under the imposed $\Z_2$ symmetry are free to mix with the usual SM particles and therefore responsible for generation of neutrino masses through type~III seesaw, explaining the observed mass-squared differences in the neutrino oscillation experiments. On the other hand, the remaining third triplet does not contribute to neutrino mass generation through seesaw, because, it is odd under imposed $\Z_2$ symmetry.  The neutral component of third triplet mix with the $\Z_2$ odd sterile neutrino $\nu_s$ to produce a 'low mass' dark matter candidate. If $\nu_s$ is light enough and its mixing with $\Sigma^0$ is small, the $\nu_s$-like mass eigenstate can be a viable dark matter candidate.

The extra renormalisable piece of Lagrangian with the triplet and the sterile neutrino is given by~\cite{Abada:2007ux, Abada:2008ea, Biggio:2011ja}
\begin{equation}
\label{Lfermtriptwobytwo}
{\cal L}=Tr [ \overline{\Sigma_a} i \slash \hspace{-2.5mm} D  \Sigma_a ] 
-\frac{1}{2} Tr [\overline{\Sigma_a}  M_\Sigma \Sigma^c_a 
                +\overline{\Sigma^c_a} M_\Sigma^* \Sigma_a] 
- \tilde{\Phi}^\dagger \overline{\Sigma_b} \sqrt{2}Y_\Sigma L 
-  \overline{L}\sqrt{2} {Y_\Sigma}^\dagger  \Sigma_b \tilde{\Phi} +\frac{i}{2} \overline{\nu_s} \slash \hspace{-2.5mm} \partial \nu_s
-\frac{1}{2} M_{\nu_s} \overline{\nu_s} \nu_s\, ,
\end{equation} 
with $L\equiv (\nu , l)^T$, $\Phi\equiv (\phi^+, \phi^0)^T\equiv
(\phi^+, (v+H+i \eta)/\sqrt{2})^T$, $\tilde \Phi = i \tau_{2} \Phi^*$,
$\Sigma^c \equiv C \overline{\Sigma}^T$ and
where summation over $a$ and $b$ are implied, with
$a = 1,2,3$, $b = 1,2$ denote generation indices for the triplets. Note that $b$ does not assume the third index as the third generation triplet is odd under $\Z_2$.

It should be noted that, in  eqn.~(\ref{Lfermtriptwobytwo}), the Yukawa coupling terms for the $\Z_2$ odd third triplet and sterile neutrino are prohibited by the exactly conserved $\Z_2$ symmetry. Also, in the covariant derivative in eqn.~(\ref{Lfermtriptwobytwo}), no $B_\mu$ terms are present, which in turn restricts the interaction between fermion triplets and $Z_\mu$ boson, this is the speciality of the presence of real triplets in the theory.

The smallness of the $\nu_s$ -- $\Sigma^0$ mixing can be generated by  dimension-five terms. In general, we add the following terms to the Lagrangian: 
\begin{equation}
{\cal L}_5= \left( {\alpha_{\Sigma \nu_s}} \Phi^{\dagger} \overline{\Sigma}  \Phi \nu_s + {\rm h.c.}\right) + {\alpha_{\nu_s}}  \Phi^{\dagger}  \Phi \overline{\nu_s}  \nu_s
+ {\alpha_{\Sigma}}  \Phi^{\dagger} \overline{\Sigma}  \Sigma \Phi  
\end{equation}
where ${\alpha_{\Sigma \nu_s}}$, ${\alpha_{\nu_s}}$  and ${\alpha_{\Sigma}}$ are three coupling constants of mass dimension $-1$.

$\nu_s$ and $\Sigma^0$ mix to produce two mass eigenstates, $\chi$
and $\Psi$, given by 
\begin{eqnarray}
\chi&=&\cos\beta \, \Sigma^0 - \sin\beta \, \nu_s \\
\Psi&=&\sin\beta \, \Sigma^0 + \cos\beta \, \nu_s 
\end{eqnarray}
We denote $\chi$ as the lighter mass eigenstate and hence, as our candidate for dark matter. The mixing angle $\beta$  is determined in terms of the five independent parameters present in the theory, given by 
\begin{equation}\label{oomagn}
M_\Sigma,\qquad M_{\nu_s},\qquad {\alpha_{\Sigma \nu_s}},\qquad {\alpha_{\Sigma}} \qquad {\rm and} \qquad {\alpha_{\nu_s}}  .  
\end{equation}

The $H\chi \chi$ vertex driving the DM annihilation in the $s$-channel, is proportional to 
\begin{equation}
\left(\sqrt{2} {\alpha_{\Sigma \nu_S}}  \cos \beta\sin\beta
 + {\alpha_{\nu_S}} \sin^2 \beta + {\frac{\alpha_{\Sigma}}{2}} \cos^2 \beta   \right)\, .
\end{equation}

 In absence of the dimension-five couplings, $\chi$ does not couple to the Higgs. Hence Higgs portal DM annihilation can not take place. As mentioned earlier, $Z$-portal DM annihilation is also not possible. In such a scenario, DM can annihilate {\it via} $\Sigma^+$ exchange in the $t$-channel to a pair of $W$ bosons. However, in this case, correct relic density can be obtained if the DM mass is larger than 2~TeV as the process is driven by unsuppressed gauge interactions. Introduction of the term ${\alpha_{\Sigma}}  \Phi^{\dagger} \overline{\Sigma}  \Sigma \Phi $ induces a  mixing between $\Sigma^0$ and $\nu_s$ which helps us to get a  $\nu_s$-like DM $\chi$. This DM then can self-annihilate {\it via} suppressed couplings with $H$ and $W$. As we are allowing this dimension-five term, for completeness we have added the other two dim-5 terms as well, which in turn give more room in the parameter space to manoeuvre.

Clearly if all the $\alpha$s are of the same order, the term ${\alpha_{\Sigma}}  \Phi^{\dagger} \overline{\Sigma}  \Sigma \Phi $ will be less effective during annihilation, because for a $\nu_s$-like dark matter $\chi$, the value of mixing coefficient $\cos\beta$ will be rather small and that will in turn suppress the effect of parameter ${\alpha_{\Sigma}}$ on the observable parameters of the model.

\section{Results}
\label{S:results}
DM mass depends mainly on the choice of $M_\Sigma$ and $M_{\nu_s}$. But it is also dependent on the higher dimensional parameters. The mass matrix for $\nu_s$ and $\Sigma^0$  is given by 
\begin{equation}
\mathcal{M}  = \left(\begin{array}{cc}
M_{\nu_s}-\alpha_{\nu_s} v^2 & \alpha_{\Sigma \nu_s} v^2 \\
\alpha_{\Sigma {\nu_s}} v^2 & M_\Sigma - \alpha_\Sigma v^2 
\end{array}\right) 
\end{equation}
The DM particle can be identified with the lighter mass eigenstate of $\mathcal{M}$.
In the limit when $M_\Sigma\simeq M_{\nu_s}$ the DM mass is a complicated function of all the five free parameters of our model. In the limit where the mixing between $\Sigma^0$ and $\nu_s$ is rather small and $M_\Sigma\gg M_{\nu_s}$, the DM mass is approximately expressed as,
\beq
M_\chi \sim M_{\nu_s}-\alpha_{\nu_s} v^2 + \frac{(\alpha_{\Sigma \nu_s} v^2)^2}{M_\Sigma}\, .
\eeq
The flexibility of our model is that, it allows `low' DM mass for a suitable choice of the relevant parameters, as demonstrated in the following. We use \texttt{FeynRules~2.0}~\cite{Alloul:2013bka}  in conjunction with \texttt{micrOmegas~3.3.13}~\cite{micrOMEGAs2, Belanger:2013oya} to compute relic density in our model.

In general, two different situation may arise, both of which contribute to the relic density, they are, (i) $M_\Sigma\gg M_{\nu_s}$ and (ii) $M_\Sigma\simeq M_{\nu_s}$. We have also noted that, in the second case, coannihilation between the charged components of the triplet $\Sigma^{\pm}$, the neutral eigenstate of higher mass $\Psi$ and the dark matter candidate $\chi$ becomes important in calculation of relic density.

If the mass difference between the DM and other $\Z_2$-odd particles are within $5$--$10$\%, it is expected that coannihilation will dominate over DM self-annihilation~\cite{griest}. Some idea about the various channels of annihilation can be found from the following benchmark values. When the mass difference between the DM candidate ($\chi$) and other $\Z_2$-odd particles is $\approx 200$~GeV with the DM mass  $\approx 875$~GeV, then the annihilation channels in decreasing order of dominance are $\chi\chi\ra H H$,  $\chi\chi\ra Z Z$ and $\chi\chi\ra W^+ W^-$. Similarly, for a the mass difference is $\approx 70$~GeV and DM mass $\approx 725$~GeV, the main annihilation channels in similar order are $\chi\chi\ra W^+ W^-$, $\chi\chi\ra H H$ and $\chi\chi\ra Z Z$. For a mass difference $\approx 30$~GeV and DM mass $\approx 770$~GeV, coannihilation takes place {\it via} $\Sigma^-\Psi\ra t b$, $\Sigma^-\Psi\ra u d$, $\Sigma^-\Psi\ra c s$, $\Sigma^-\Sigma^-\ra W^- W^-$ and $\Psi\Psi\ra W^+ W^-$. For mass difference $\approx 700$~GeV ($1.3$~TeV) and DM mass $\approx 230$~GeV ($197$~GeV), annihilation proceeds along $\chi\chi\ra H H$. Finally, for mass difference $\approx 550$~GeV with DM mass $\approx 425$~GeV, annihilation through both the channels $\chi\chi\ra H H$ and $\chi\chi\ra W^+ W^-$ becomes important.

\begin{figure}
\centerline{
        \includegraphics[width=0.75\textwidth]{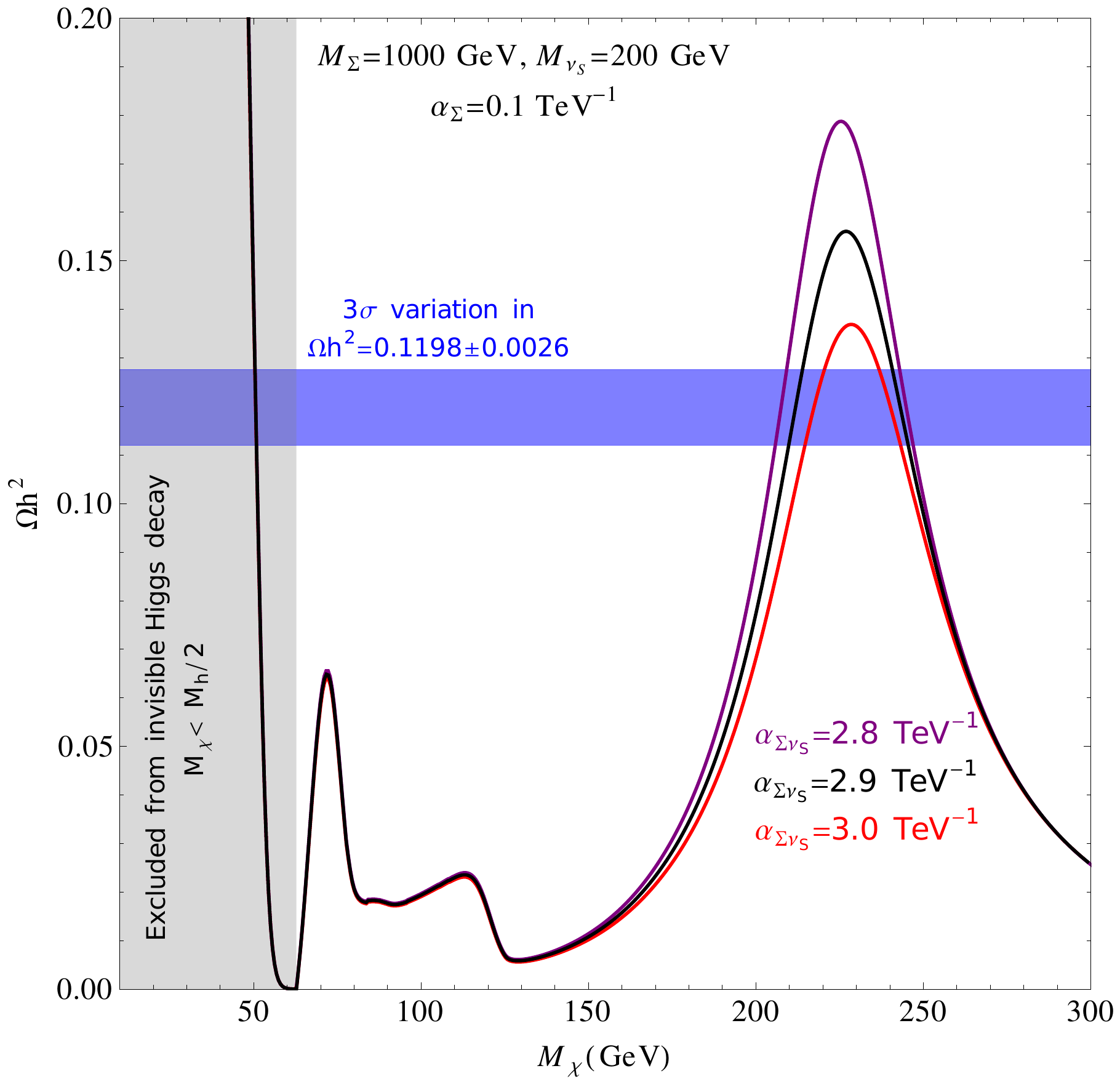}
}
\caption{\textit {Relic density $\Omega h^2$ vs. dark matter mass $M_{\chi}$ keeping $\alpha_{\Sigma\nu_s}$ fixed at $2.8$~TeV$^{-1}$, $2.9$~TeV$^{-1}$ and $3.0$~TeV$^{-1}$. $M_{\chi}$ is varied by changing the remaining parameter $\alpha_{\nu_s}$. We use $M_\Sigma = 1$~TeV, $\alpha_\Sigma = 0.1$~TeV$^{-1}$, $M_{\nu_s} = 200$~GeV. The blue band corresponds to $3 \sigma$ variation in relic density  according to the WMAP + Planck data.} \label{figsr1}}
\end{figure}

In Fig.~\ref{figsr1}, we fix $M_\Sigma = 1$~TeV and $\alpha_\Sigma = 0.1$~TeV$^{-1}$. We also choose a benchmark value of $M_{\nu_s} = 200$~GeV. We plot DM relic density against the DM mass $M_\chi$. DM mass is varied
  by changing  $\alpha_{\nu_s}$, but keeping $\alpha_{\Sigma\nu_s}$ fixed at different  values. The lightly shaded region indicates the region that is excluded when $M_{\chi} < M_h/2$, where $M_h$ is the Higgs mass, due to the constraint on invisible Higgs decay width. The WMAP~\cite{wmap} and Planck~\cite{Ade:2013zuv} allowed DM relic density being too restrictive, we see from the figure, narrow regions of DM mass $\sim 210$~GeV and $\sim 240$~GeV are allowed. However, for this specific choice of parameters in this plot, only DM mass $\sim 210$~GeV is allowed from LUX direct detection constraints.
 
\begin{figure}
    \centering
     \subfloat[]{\includegraphics[width=0.45\textwidth]{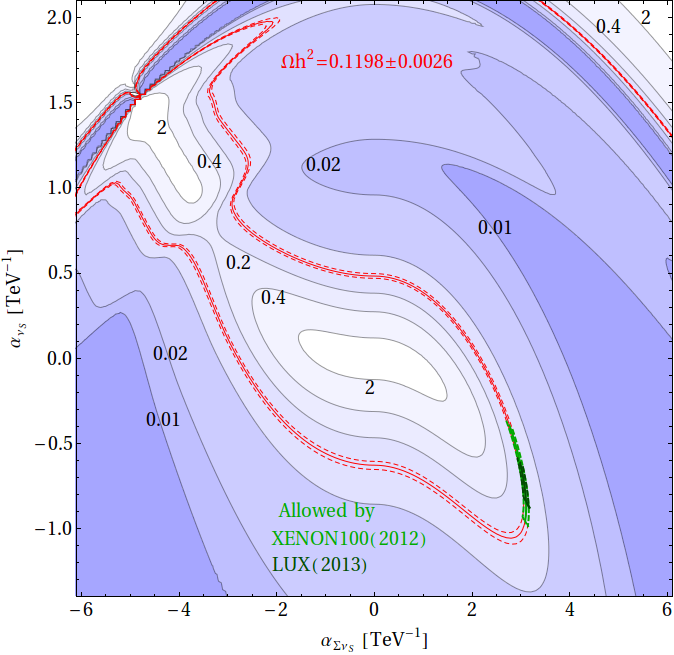}\label{fig1a}}
     \subfloat[]{\includegraphics[width=0.45\textwidth]{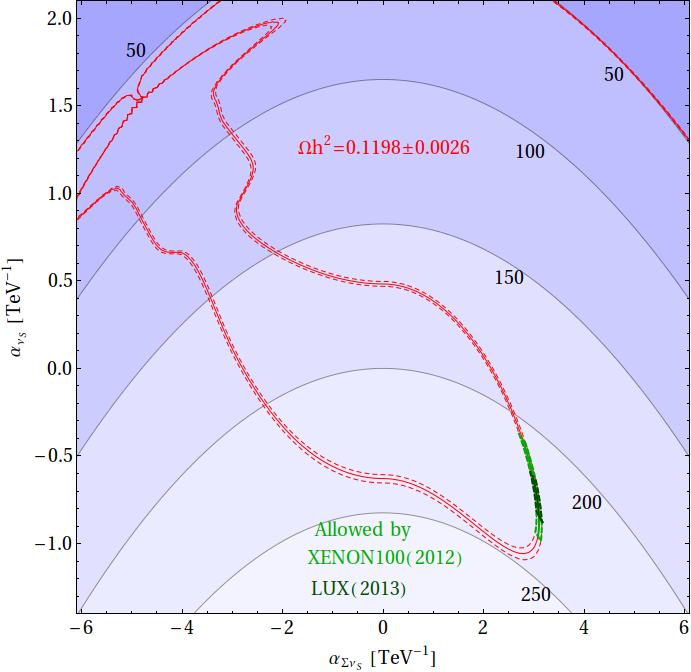}\label{fig1b}}
     \caption{\textit{Contour plot in $\alpha_{\Sigma \nu_s}$ -- $\alpha_{\nu_s}$ plane for fixed values of $M_\Sigma = 1$~TeV, $M_{\nu_s} = 200$~GeV,  $\alpha_\Sigma = 0.1$~TeV$^{-1}$. In {\rm (a)} DM relic density is projected onto the plane, whereas in {\rm (b)}, we project DM mass (in GeV). In both plots the red solid line represents $\Omega h^2 = 0.1198$ and the dashed red lines correspond to the $3 \sigma$ variation in $\Omega h^2$. Darker region corresponds to lower values of $\Omega h^2$ or $M_\chi$.}}
     \label{figsr2}
\end{figure}

 In order to find the favoured parameter space which satisfies DM relic density constraints, in Fig.~\ref{figsr2}(a) we present contour plots of $\Omega h^2$ in the  $\alpha_{\Sigma \nu_s}$ -- $\alpha_{\nu_s}$ plane for $M_\Sigma = 1000$~GeV and $M_{\nu_s}= 200$~GeV. The solid red curve denotes $\Omega h^2=0.1198$. A $3\sigma$ error band is shown by the dashed red curves. We have varied $\alpha_{\Sigma \nu_s}$ from $-6.0$ to $6.0$~TeV$^{-1}$ and $\alpha_{\nu_s}$ from $-1.5$ to $2.0$~TeV$^{-1}$ to obtain this plot. We have also fixed $\alpha_\Sigma = 0.1$~TeV$^{-1}$ for this plot. In Fig.~\ref{figsr2}(b) we project DM mass  $M_\chi$ onto the  $\alpha_{\Sigma \nu_s}$ -- $\alpha_{\nu_s}$ plane for same values of parameters used in the previous figure. 
The light green region and the deep green region in both plots corresponding to $M_\chi$ between $205 - 235$~GeV
 indicate the parameter space which satisfy DM relic density constraints as well as WIMP-nucleon cross section bound as imposed by XENON~100~\cite{xenon2011, xenon2012} and LUX~\cite{Akerib:2013tjd}. 

\begin{figure}
     \centering
     \subfloat[]{\includegraphics[width=0.45\textwidth]{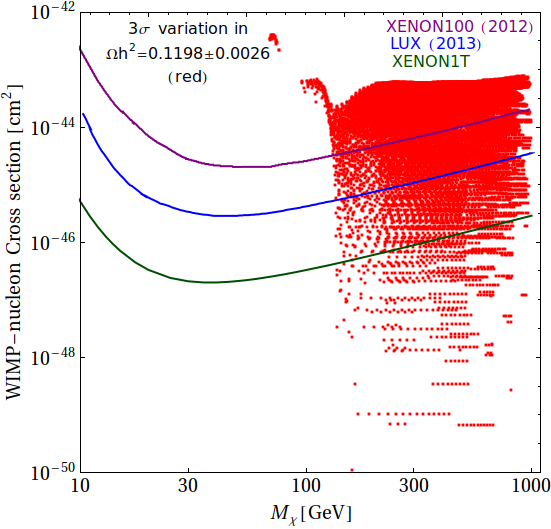}\label{fig2a}}
     \subfloat[]{\includegraphics[width=0.45\textwidth]{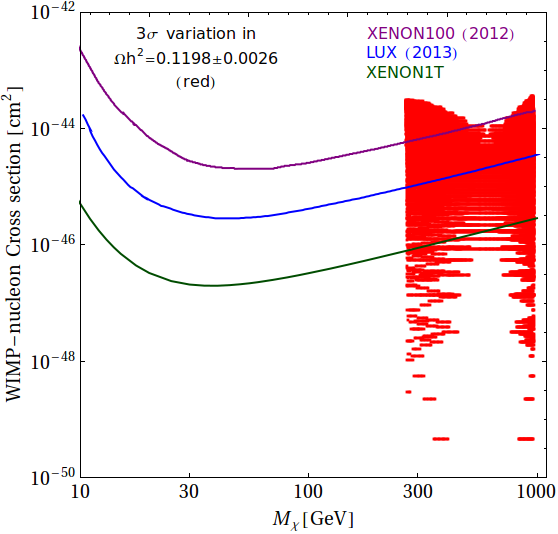}\label{fig2b}}
     \caption{\textit {{\rm (a)} WIMP-nucleon cross section vs. dark matter mass $M_{\chi}$  keeping $\alpha_\Sigma$ fixed at $0.1$ TeV$^{-1}$. $M_{\chi}$ is varied by changing the remaining parameters. In such a scenario, DM annihilates {\it via} self-annihilation only. $M_{\nu_s}$ is varied between $150$ and $1000$~GeV. 
     {\rm (b)} WIMP-nucleon cross section vs. dark matter mass $M_{\chi}$  keeping $\alpha_\Sigma$ fixed at $0.1$ TeV$^{-1}$. $M_{\chi}$ is varied by changing the remaining parameters. Here coannihilation between the DM candidate with the charged components of the triplet $\Sigma^{\pm}$, the neutral eigenstate of higher mass provides the dominating contribution to relic density. In both the figures different direct detection experimental bounds are indicated and the red points correspond to the $3 \sigma$ variation in relic density according to the WMAP + Planck data.  }}
     \label{figsr3}
\end{figure}


\begin{figure}
 \centering
 \subfloat[]{\includegraphics[angle=-90,width=0.45\textwidth]{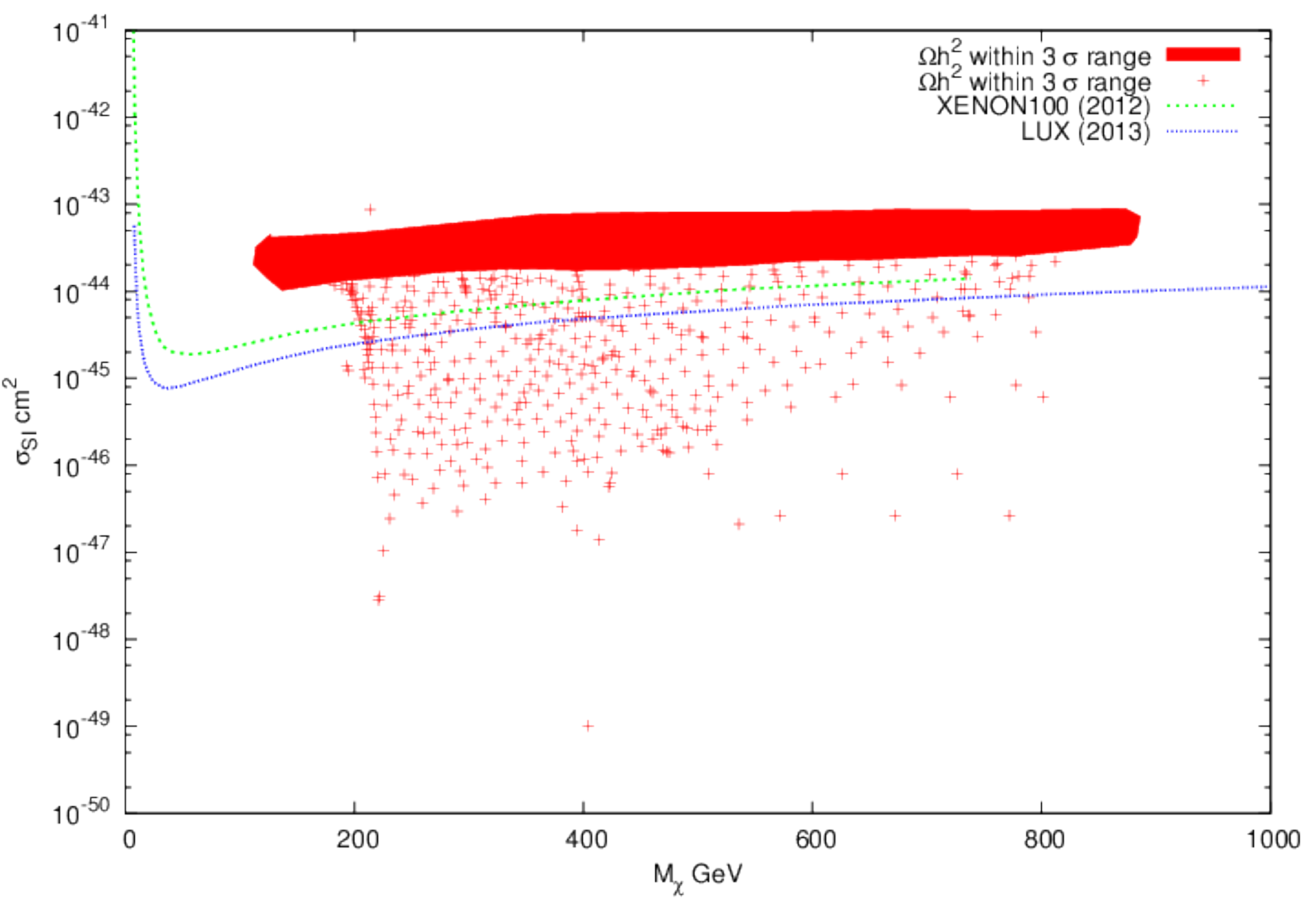}\label{fig4a}}
  \subfloat[]{\includegraphics[angle=-90,width=0.45\textwidth]{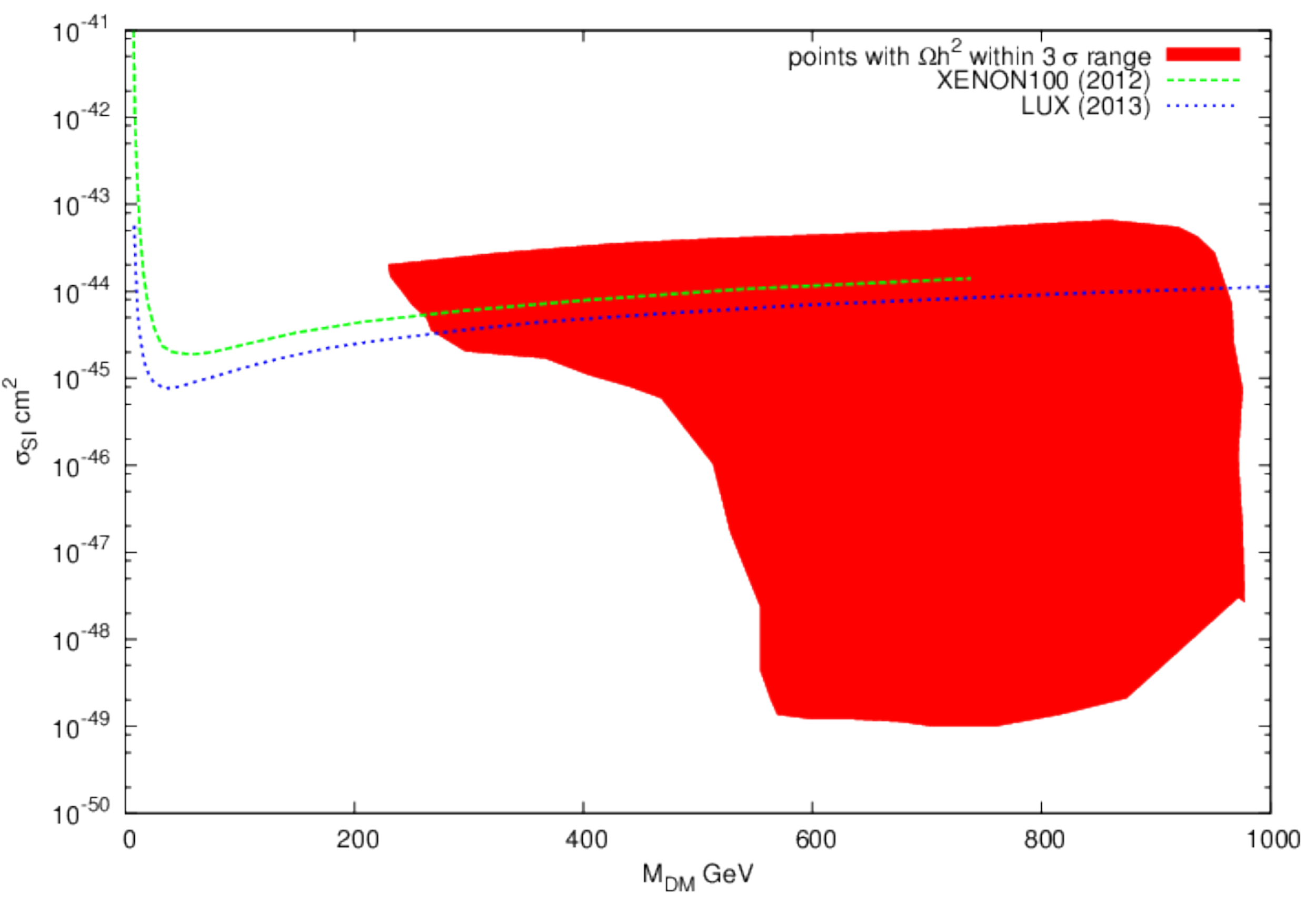}\label{fig4b}}
     \caption{\textit {{\rm (a)} WIMP-nucleon cross section vs. dark matter mass $M_{\chi}$  keeping $\alpha_\Sigma$ fixed at $0.01$ TeV$^{-1}$. $M_{\chi}$ is varied by changing the remaining parameters. Here self-annihilation of the DM candidate provides the dominating contribution to relic density
     {\rm (b)} WIMP-nucleon cross section vs. dark matter mass $M_{\chi}$  keeping $\alpha_\Sigma$ fixed at $0.01$ TeV$^{-1}$. $M_{\chi}$ is varied by changing the remaining parameters. Here coannihilation between the DM candidate with the charged components of the triplet $\Sigma^{\pm}$, the neutral eigenstate of higher mass provides the dominating contribution to relic density. In both the figures different direct detection experimental bounds are indicated and the red points correspond to the $3 \sigma$ variation in relic density according to the WMAP + Planck data. }}
     \label{figsr4}
\end{figure}

 In Fig.~\ref{figsr3}(a) we have plotted the WIMP-nucleon cross section vs. dark matter mass $M_{\chi}$  keeping $\alpha_\Sigma$ fixed at $0.1$ TeV$^{-1}$. $M_{\chi}$ is varied by changing the remaining parameters. For illustration, we have varied $M_\Sigma$ between $300$ and $1000$~GeV, $M_{\nu_s}$ between $150$ and $1000$~GeV keeping the mass difference between DM and other $\Z_2$ odd fermions greater than $200$~GeV. The remaining parameters $\alpha_{\Sigma \nu_s}$ and $\alpha_{\nu_s}$ were varied between $-8$~TeV$^{-1}$ to $8$~TeV$^{-1}$ and $-0.9$~TeV$^{-1}$ to $0.9$~TeV$^{-1}$ respectively. In this case, the DM particle always self-annihilates and DM coannihilation does not take place. It is seen that a large region of parameter space satisfies the bound on WIMP-nucleon cross section as imposed by XENON~100~\cite{xenon2011, xenon2012} and LUX~\cite{Akerib:2013tjd} experimental data. In Fig.~\ref{figsr3}(b) we have plotted the WIMP-nucleon cross section vs. dark matter mass $M_{\chi}$  keeping $\alpha_\Sigma$ fixed at $0.1$~TeV$^{-1}$. We varied $M_\Sigma$ between $300$ and $1000$~GeV, $M_{\nu_s}$ between $270$ and $1000$~GeV keeping the mass difference between DM mass and other $\Z_2$ odd fermions always smaller than $30$~GeV. The other parameters $\alpha_{\Sigma \nu_s}$ and $\alpha_{\nu_s}$ were varied between $-8$~TeV$^{-1}$ to $8$~TeV$^{-1}$ and $-0.9$~TeV$^{-1}$ to $0.9$~TeV$^{-1}$ respectively. In this case, we noted that coannihilation between the charged components of the triplet $\Sigma^{\pm}$, the neutral eigenstate of higher mass ($\Psi$) and the dark matter candidate ($\chi$) provides the dominant contributions in the calculation of relic density. Again a large region of parameter space satisfies the bound on WIMP-nucleon cross section as imposed by XENON~100 and LUX experimental data~\cite{dmtools} in this case too.
 
 In Fig.~\ref{figsr4}(a), we have plotted the WIMP-nucleon cross section vs. dark matter mass $M_{\chi}$  keeping $\alpha_\Sigma$ fixed at $0.01$ TeV$^{-1}$. Other parameters $\alpha_{\Sigma \nu_s}$ and $\alpha_{\nu_s}$ were varied between $1.2$~TeV$^{-1}$ to $3.0$~TeV$^{-1}$ and $-1.1$~TeV$^{-1}$ to $0.03$~TeV$^{-1}$ respectively. In this case, we varied  $M_\Sigma$ between $500$ and $1500$~GeV and $M_{\nu_s}$ between $150$ and $900$~GeV. We obtained wide region of DM mass from $128$~GeV to $872$~GeV that satisfies the correct relic density range and the constrains imposed on WIMP-nucleon cross section, i.e., $\sigma_{SI}$ by XENON~100 and LUX data. The mass difference between  $M_\Sigma$ and  $M_{\nu_s}$ was always kept greater than $150$~GeV. In this case also the annihilation of DM particle with itself provides the dominating contribution to the calculation of relic density. In Fig.~\ref{figsr4}(b), we have plotted the WIMP-nucleon cross section vs. dark matter mass $M_{\chi}$  keeping $\alpha_\Sigma$ fixed at $0.01$ TeV$^{-1}$. Other parameters $\alpha_{\Sigma \nu_s}$ and $\alpha_{\nu_s}$ were varied between $-0.08$~TeV$^{-1}$ to $0.94$~TeV$^{-1}$ and $-0.009$~TeV$^{-1}$ to $0.096$~TeV$^{-1}$ respectively. In this case we varied  $M_\Sigma$ between $300$ and $1000$~GeV and $M_{\nu_s}$ in such a way as to make the mass difference between $M_\Sigma$ and  $M_{\nu_s}$ smaller than $55$~GeV. In this case also coannihilation process between the charged components of the triplet $\Sigma^{\pm}$, the neutral eigenstate of higher mass ($\Psi$) and the dark matter candidate ($\chi$) provides the dominant contributions in the calculation of relic density. Here also we got a large region of parameter space that satisfies the correct relic density requirement and the bound imposed by XENON~100 and LUX data on WIMP-nucleon cross section. In both the plots the red points and red regions correspond to those data points that fall within the $3 \sigma$ variation in relic density according to the combined WMAP and Planck data. 

\section{Summary and Conclusion}
\label{S:conclusion}
To explain smallness of neutrino masses, type-III seesaw mechanism employs heavy fermion triplets. In this paper, we have tried to use such a scenario to also address the dark matter riddle.  Previous attempts in this regard employed radiative mass splitting of the members of a lepton triplet, yielding DM masses at the TeV scale. We have in addition introduced a sterile neutrino to bring down the DM mass scale to $\sim 100$~GeV. Both the additional triplet and the sterile neutrino are odd under a postulated $\Z_2$ symmetry.

Dark matter direct detection experimental results are rather conflicting. While CDMS~\cite{cdms_science, Agnese:2013cvt}, DAMA~\cite{dama}, CoGeNT~\cite{cogent}, CRESST~\cite{cresst} {\it etc.} point towards a DM mass of $\sim 10$~GeV, XENON~100 and LUX claim to rule out these observations. Given this scenario, we keep our options open and explore all DM masses allowed by experimental observations. 

As we are exploring a Higgs portal dark matter model, to respect the invisible Higgs decay width constraints we consider DM masses more than $M_h/2\sim 62.5$~GeV. We have ensured that our results are consistent with XENON~100 and LUX constraints. In addition, we also indicate the constraints on the parameters of our model ensuing from WMAP and Planck data. In our model, the charged triplet fermions, being $\Z_2$ odd, do not couple to SM fermions  
and as a consequence, can evade detection in the existing colliders including LHC.

We have introduced dimension five terms to satisfy the DM relic density constraints with the help of a sterile neutrino-like DM. The role of these terms are to produce the right amount of $\Sigma^0$--$\nu_s$ mixing to get the appropriate DM mass and DM annihilation. 

It is possible to satisfy the relic density constraints in a model with a pure sterile neutrino DM with the dimension five  term ${\alpha_{\nu_s}}  \Phi^{\dagger}  \Phi \overline{\nu_s}  \nu_s$ alone, without the triplet fermions. However in this model the allowed parameter space is rather restricted compared to our model which offers a lot of flexibility in DM mass.

In short, we have presented a triplet fermion DM model, which can provide us with DM candidates with masses as low as $\approx 100$~GeV. The scenario, however has the flexibility to accommodate a higher mass DM candidate as well, and offers a wide region of parameter space consistent with observations

\section{Acknowledgements}
N.K. acknowledges financial support from University Grants Commission, India. 
A.C and B.M. acknowledges the funding available from the Department of Atomic Energy, Government of India, for the Regional Centre for Accelerator based Particle Physics~(RECAPP), Harish-Chandra Research Institute. 
S.R. acknowledges support of seed grant from Indian Institute of Technology Indore, while A.C. thanks Indian Institute of Technology Indore for hospitality during the work. Both N.K. and S.R. are grateful to RECAPP for hospitality while the work was on progress. 
N.K. is thankful to Kamakshya P. Modak and Danielle H. Speller for useful communications. A.C. thanks Subhadeep Mondal and Tanumoy Mondal for useful discussions.

\end{document}